# Magnetization-tuned topological quantum phase transition in $MnBi_2Te_4$ devices

Jun Ge[1], Yanzhao Liu[1], Pinyuan Wang[1], Zhiming Xu[2], Jiaheng Li[2], Hao Li[3,4], Zihan Yan[1], Yang Wu[4,5], Yong Xu[2,6], Jian Wang[1,2,7,8*]

[1] International Center for Quantum Materials, School of Physics, Peking University, Beijing 100871, China

[2] State Key Laboratory of Low Dimensional Quantum Physics, Department of Physics, Tsinghua University, Beijing 100084, China

[3] School of Materials Science and Engineering, Tsinghua University, Beijing 100084, China

[4] Tsinghua-Foxconn Nanotechnology Research Center and Department of Physics, Tsinghua University, Beijing 100084, China

[5] Department of Mechanical Engineering, Tsinghua University, Beijing 100084, China

[6] RIKEN Center for Emergent Matter Science (CEMS), Wako, Saitama 351-0198, Japan

[7] CAS Center for Excellence in Topological Quantum Computation, University of Chinese Academy of Sciences, Beijing 100190, China

[8] Beijing Academy of Quantum Information Sciences, Beijing 100193, China

Recently, the intrinsic magnetic topological insulator $MnBi_2Te_4$ has attracted enormous research interest due to the great success in realizing exotic topological quantum states, such as the quantum anomalous Hall effect (QAHE), axion insulator state, high-Chern-number and high-temperature Chern insulator states. One key issue in this field is to effectively manipulate these states and control topological phase transitions. Here, by systematic angle-dependent transport measurements, we reveal a magnetization-tuned topological quantum phase transition from Chern insulator to magnetic insulator with gapped Dirac surface states in $MnBi_2Te_4$ devices. Specifically, as the magnetic field is tilted away from the out-of-plane direction by around 40-60 degrees, the Hall resistance deviates from the quantization value and a colossal, anisotropic magnetoresistance is detected. The theoretical analyses based on modified Landauer-Büttiker formalism show that the field-tilt-driven switching from ferromagnetic state to canted antiferromagnetic state induces a topological quantum phase transition from Chern insulator to magnetic insulator with gapped Dirac surface states in $MnBi_2Te_4$ devices. Our work provides an efficient means for modulating topological quantum states and topological quantum phase transitions.

Topological quantum states, such as quantum anomalous Hall effect (QAHE) [1, 2] and axion insulator states [3], have potential in generating new device concepts for topological quantum computing, low-power-consumption electronics and spintronics physics [4-7]. The recently discovered intrinsic magnetic topological insulator (TI) $MnBi_2Te_4$ provides an ideal platform for exploring exotic topological quantum states and has become a focus of attention of condensed matter physics. $MnBi_2Te_4$ is a layered magnetic topological material [8-24]. As shown in Fig. 1(a), the monolayer $MnBi_2Te_4$ is formed by a Te-Bi-Te-Mn-Te-Bi-Te septuple layer (SL). In each SL, the $Mn^{2+}$ ions contribute the magnetic moments and the intralayer exchange coupling results in the ferromagnetic (FM) order with an out-of-plane easy axis [8]. Between neighboring SLs, the weak interlayer antiferromagnetic (AFM) exchange coupling generates the A-type AFM order [8]. Due to its unique layered structures and weak interlayer AFM coupling, abundant topological phases can be expected in different magnetic states of $MnBi_2Te_4$ [8, 9, 23-25]. For example, in the absence of magnetic field, bulk $MnBi_2Te_4$ has been experimentally verified to be an AFM TI [10, 11]. When bulk $MnBi_2Te_4$ is exfoliated into thin flakes, many topological phases can be detected by controlling the layer number [8, 9, 23]. In even-layer $MnBi_2Te_4$ thin flakes, the magnetic moments are completely compensated because of the interlayer AFM coupling [8, 23]. The requirements for realizing axion insulator state may thus be fulfilled [8, 26] and recently the signature of axion insulator state has been detected in 6-SL $MnBi_2Te_4$ devices [13]. Odd-layer $MnBi_2Te_4$ thin flakes with three or more layers is able to exhibit net magnetization due to the uncompensated antiferromagnetism [8, 23] and QAHE is theoretically predicted [8] and has been experimentally observed in one 5-SL $MnBi_2Te_4$ thin flake [12]. When driven into ferromagnetic (FM) state by external magnetic field, bulk $MnBi_2Te_4$ turns into the simplest magnetic Weyl semimetal (WSM) with only one pair of Weyl points [8, 9, 14, 24, 27]. Further fabricating $MnBi_2Te_4$ into thin devices, high-Chern-number Chern insulator (CI) states have been experimentally demonstrated in the Weyl semimetal phase of $MnBi_2Te_4$ [14]. For both topological physics and related applications, effectively manipulating these states and controlling topological phase transitions are highly desired.

In this Letter, we reveal a magnetization-tuned topological quantum phase transition from CI to magnetic insulator with gapped Dirac surface states in $MnBi_2Te_4$ devices by systematic angle-dependent transport measurements. We show that Chern insulator states with Chern number $C$=2 and $C$=1 can be realized in 9-SL and 8-SL devices by applying perpendicular magnetic field, respectively. As the magnetic field is tilted away from the perpendicular direction, a switch from edge state-dominated dissipationless transport to diffusive transport is detected. Further analysis based on modified Landauer–Büttiker formalism shows that the magnetic structure transition from FM state to canted antiferromagnetic (cAFM) state leads to a topological quantum transition from CI to magnetic insulator with gapped Dirac surface states. Our work provides an efficient method for modulating topological quantum phase transitions, and the observation of the related colossal anisotropic longitudinal magnetoresistance shows the potential in electronics and spintronics applications.

Due to the weak van der Waals force between neighboring SLs, $MnBi_2Te_4$ thin flakes with different thicknesses are easily obtained through mechanical exfoliation method [12-14]. These flakes were then transferred onto the $SiO_2$/Si substrate and fabricated into devices with Hall bar geometry (Fig. 1(b)). In device D1 (9-SL), quantized Hall resistance ($R_{yx}$) plateau with height of 0.986 $h/2e^2$ is observed (Fig. 1(c)), accompanied by a longitudinal resistance ($R_{xx}$) as small as 0.021 $h/2e^2$ (Fig. 1(d)), indicating the observation of high-Chern-number Chern insulator state with two sets of dissipationless chiral edge states. In device D2 (8-SL), the Chern insulator state with $C$=1 is detected, characterized by well quantized Hall resistance plateau ($h/e^2$) and nearly vanishing $R_{xx}$. As shown in our previous report, the Chern insulator state in $MnBi_2Te_4$ devices originates from the magnetic WSM phase in FM state [14]. An important issue thus remains to be explored in this context: how will the Chern insulator states evolve as the magnetic structure changes?

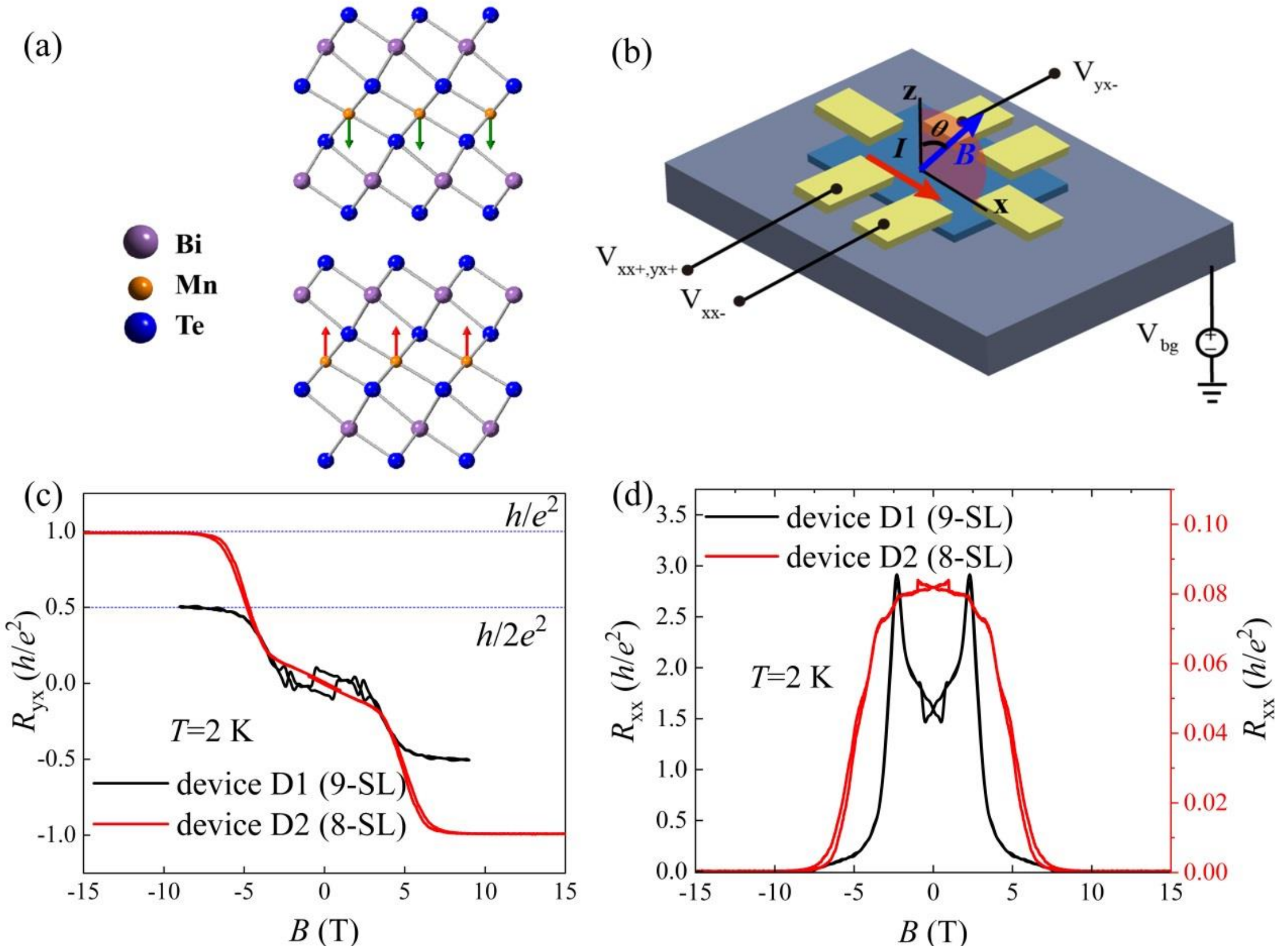

FIG. 1. (a) Schematic crystal and magnetic structure of $MnBi_2Te_4$. The orange and cyan arrows denote magnetic moment directions of Mn ions. (b) Schematic diagram

of the transport measurements configuration. The magnetic field rotates from out-of-plane direction to in-plane direction and the angle between magnetic field and the out-of-plane direction is denoted as $\theta$. (c), (d) $R_{yx}$ and $R_{xx}$ of device D1 (9-SL) and D2 (8-SL) as a function of magnetic field. The dashed lines in (c) denote the quantized values $h/e^2$ and $h/2e^2$, respectively.

To explore the relation between magnetic structures and topological phases, we carry out transport measurements at different magnetic structures by tilting the magnetic field. We firstly perform angle-dependent transport measurements on high-quality $MnBi_2Te_4$ bulk single crystals at 2 K to examine how the magnetic field influences magnetic structures. Figure 2(a) shows the schematic plot of the angle-dependent transport measurement configuration. The external magnetic field rotates from the out-of-plane direction to the in-plane direction. The tilted angle $\theta$ of the magnetic field with respect to out-of-plane direction is varied from 0°(out-of-plane) to 90°(in-plane). Figures 2(b) and 2(c) exhibit the Hall resistivity $\rho_{yx}$ and longitudinal resistivity $\rho_{xx}$ as a function of magnetic field $B$ at various $\theta$, respectively. For clarity, data curves in Figs. 2(b) and 2(c) are shifted. At $\theta$=0°, two transitions marked by $B_1$ (~ -3 T) and $B_2$ (~ -7 T) can be clearly observed on both $\rho_{yx}$ and $\rho_{xx}$, which is consistent with previous reports [13, 28]. At $B$=$B_1$, $\rho_{yx}(B)$ shows an evident kink, simultaneously $\rho_{xx}(B)$ exhibits obvious decrease. This indicates that spins in $MnBi_2Te_4$ start to flip and the system enters a metastable cAFM state [13, 28]. Further increasing the magnetic field induces the second transition marked by $B_2$ and drives the system into FM state [13, 28]. Remarkably, this magnetic transition results in the magnetic WSM phase in the bulk [8, 24, 27]. With increasing $\theta$, $B_2$ becomes larger and when $\theta$ is above 30°, the kink on both $\rho_{yx}(B)$ and $\rho_{xx}(B)$ marked by $B_2$ cannot be detected within 9 T. Meanwhile, the steep drop on $\rho_{xx}(B)$ decreases as increasing $\theta$ and finally disappears above 50°, indicating the spin-flip process is suppressed [28]. This behavior is further confirmed by the angle-dependent magnetization ($M$) measurements results. As shown in Fig. 2(d), when $\theta \leqslant 45°$, the spin-flip-induced nonlinear $M$-$B$ behavior is observed. Further increasing $\theta$, $M$ turns to be linear with $B$ and the spin-flip process is suppressed below 7 T. These results show that tilting the magnetic field will induce the transitions between different magnetic structures when the strength of magnetic field is fixed. Thus, as the magnetic structure is closely related to the topological phases of $MnBi_2Te_4$ [8, 9, 23], the evolution of the topological quantum states can be well studied by tilting the magnetic field.

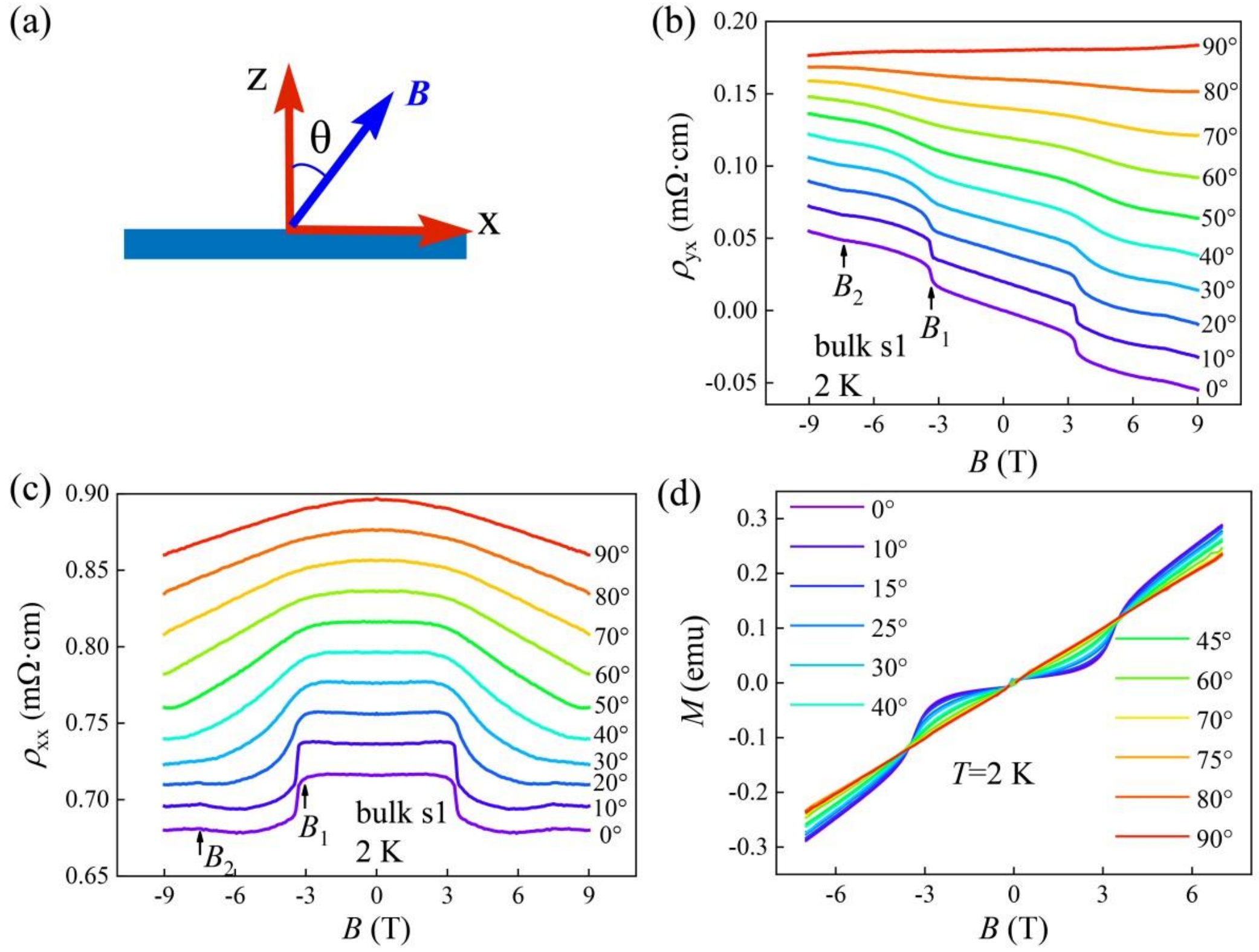

FIG. 2. (a) A schematic of the configuration highlighting field rotation from out-of-plane to in-plane. (b), (c) $\rho_{yx}$ and $\rho_{xx}$ of the $MnBi_2Te_4$ single crystal as a function of magnetic field at various angles. $B_1$ and $B_2$ represent the transition from AFM state to cAFM state, and transition from cAFM state to FM state, respectively. For clarity, data curves are shifted. (d) Angle-dependent $M(B)$ curves measured at 2 K.

Having established the filed-tilt-driven magnetic structure transitions, we now examine the evolution of Chern insulator states as the magnetization tilts away from out-of-plane direction. As shown in Figs. 3(a)-(c), the magnetic fields required to induce cAFM and FM states increase with $\theta$ in device D1 when $\theta \leq 40°$, where the quantized Hall resistance plateau and nearly vanishing longitudinal resistance can still be observed. When $\theta$ is above 40°, $R_{yx}$ deviates from quantization. Meanwhile, $R_{xx}$ at -9 T sharply increases (Fig. 3(b) and Fig. 4), indicating the switch from edge state dominated transport to diffusive transport. When $\theta$=90°, $R_{yx}$ is nearly zero and $R_{xx}$ reaches the maximum value. A colossal, anisotropic magnetoresistance (~ 43700%), defined as ($R_{xx,max}$ - $R_{xx,min}$) ×100% / $R_{xx,min}$ is revealed. The similar angle-dependent behavior can also be observed in the 8-SL device D1 (Fig. S2). The $B$-$\theta$ phase diagrams in Fig. 3(c) clearly exhibit the evolution of the high-Chern-number Chern insulator state in D1. The phase diagram is characterized by phase boundaries, $B_{AFM}$ ($\theta$) and $B_{FM}$ ($\theta$). The $B_{AFM}$ ($\theta$) data points, as the boundary of the AFM states, are composed of the peak values of the $R_{xx}$ ($B$) curves at various $\theta$ (Fig. 3(b)). The $B_{FM}$ ($\theta$) curves, as the boundaries of the FM states and the Chern insulator state, represent the magnetic fields required to realize the quantized Hall resistance plateau at different $\theta$. The cAFM state lies between $B_{AFM}$ ($\theta$) and $B_{FM}$ ($\theta$) curves. Figures 3(d)-(f) show the schematics of these magnetic structures. It is obvious that when $\theta$ is above 40°, a magnetic structure transition from FM state to cAFM state occurs at ± 9 T and the Chern insulator state is destroyed.

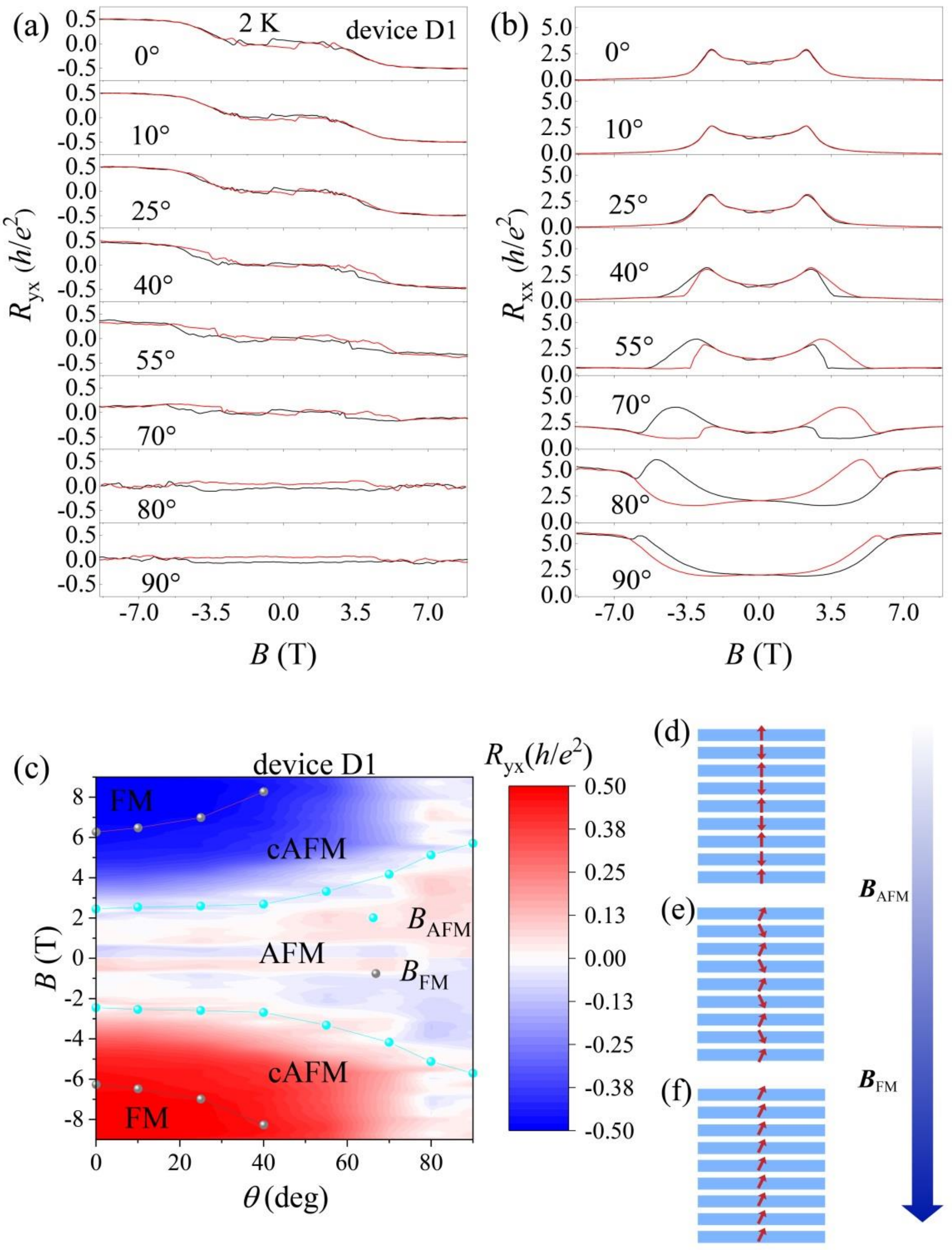

FIG. 3. (a), (b) $R_{yx}$ and $R_{xx}$ as a function of $B$ at various angles in $MnBi_2Te_4$ device D1 (9-SL). The black and red traces represent magnetic field sweeping to the negative and positive directions, respectively. (c) $B$-$\theta$ phase diagram of D1. (d-f) The schematic of the magnetic structure at different magnetic fields. When $B$ is lower than the spin-flip field $B_{AFM}$, $MnBi_2Te_4$ is in AFM state. $MnBi_2Te_4$ will be in cAFM state when $B_{AFM} < B < B_{FM}$ and finally enters the FM state when further increasing the magnetic field above $B_{FM}$.

To have a better understanding on the evolution of the Chern insulator state, we further analyze the $R_{xx}$-$\theta$ curve. As shown in Fig. 4, the $\cos^2\theta$ angular dependence commonly seen in conventional ferromagnets [29] cannot fit the angular dependence in $MnBi_2Te_4$. To explain our observations, we established a model based on the four-terminal Landauer–Büttiker formalism [30]. The well-known Landauer-Büttiker formalism can be expressed as $I_i = \sum_j G_{ij}(V_i - V_j)$, wherein $I_i$ and $V_i$ are the current and voltage of the $i_{th}$ contact, and $G_{ij} = \frac{e^2}{h} T_{ij}$ is the conductance between the $i_{th}$ and $j_{th}$ contacts [30]. Our observation of the switch from edge-state dominated ballistic transport to diffusive transport may suggest the coexistence of dissipationless edge channels and dissipative surface channels, similar as in magnetically doped topological insulators [31, 32]. Thus, the transmission coefficient $T_{ij}$ takes the form $T_{ij} = \eta\delta_{i,j-1} + t_{ij}$. Here, $\delta_{i,j-1}$ is the Kronecker delta function, which is equal to 1 if $i$=$j$-1 and 0 otherwise, $\eta$ is the transmission coefficient of the unidirectional edge modes flowing from the $i_{th}$ to the $(i+1)_{th}$ contacts, and $t_{ij}$ represents the contribution of dissipative channels. Based on these premises, we derive the following expression for $R_{xx}$ (see more details in the Supplemental Material):

$$R_{xx} = \frac{h}{e^2}\frac{t}{(\eta+t)^2}, \qquad (1)$$

where $t$ represents the dissipative contribution.

In the paramagnetic region, the three-dimensional (3D) $MnBi_2Te_4$ bulk is a time-reversal invariant TI and displays topological gapless Dirac states on the two-dimensional (2D) surfaces [8, 10]. In the cAFM region, the $MnBi_2Te_4$ bulk is a band-inverted magnetic insulator, in which the Dirac surface states persist but are gapped by the magnetism [33]. In $MnBi_2Te_4$ devices, the gapped Dirac surface states on the top and bottom are essentially decoupled from each other except for ultrathin cases (below around 4 SLs) [8, 34]. In this case, the diffusive contribution from surface channels must be considered and transmission coefficient of the unidirectional edge mode $\eta$ is suppressed from the intermixing between the edge channels and the dissipative surface channels [31]. A phenomenological expression can be used to describe this suppression as: $\eta = 1 - e^{-\frac{\Delta}{\Delta_1}}$, wherein $\Delta$ is the angle-dependent gap of the top surface states of cAFM $MnBi_2Te_4$ and $\Delta_1$ is the energy scale of the intermixing process. We further assume that the dissipative transport is dominated by the thermal excitation of surface states and takes the form $t = t_0 e^{-\frac{\Delta}{k_B T}}$. The above assumptions satisfy the physical requirement at the two extrema: $\Delta \to 0: \eta \to 0$; $\Delta \to \infty: \eta \to 1, t \to 0$. After substituting the angular dependence of the surface gap: $\Delta = \Delta_{\mathrm{m}} cos(\theta)$, wherein $\Delta_{\mathrm{m}}$ is the magnetic gap at $\theta$=0°, the angular dependence of $R_{xx}$ then can be obtained from equation (1):

$$R_{xx} = \frac{h}{e^2} \frac{t_0 e^{-\Delta_2' cos(\theta)}}{\left(1 - e^{-\Delta_1' cos(\theta)} + t_0 e^{-\Delta_2' cos(\theta)}\right)^2}. \qquad (2)$$

wherein $\Delta_1' = \frac{\Delta_{\mathrm{m}}}{\Delta_1}$ and $\Delta_2' = \frac{\Delta_{\mathrm{m}}}{k_B T}$ . We now use equation (2) to fit the $R_{xx}$-$\theta$ curve in Fig. 4. The fitting is well for $52° < \theta < 144°$, indicating that $MnBi_2Te_4$ is tuned into the magnetic insulator with gapped Dirac surface states and the dissipation process are dominated by the surface states. On the other hand, the fitting shows a deviation in the CI regime, further indicating that the Chern insulator states in $MnBi_2Te_4$ originate from the magnetic WSM phase.

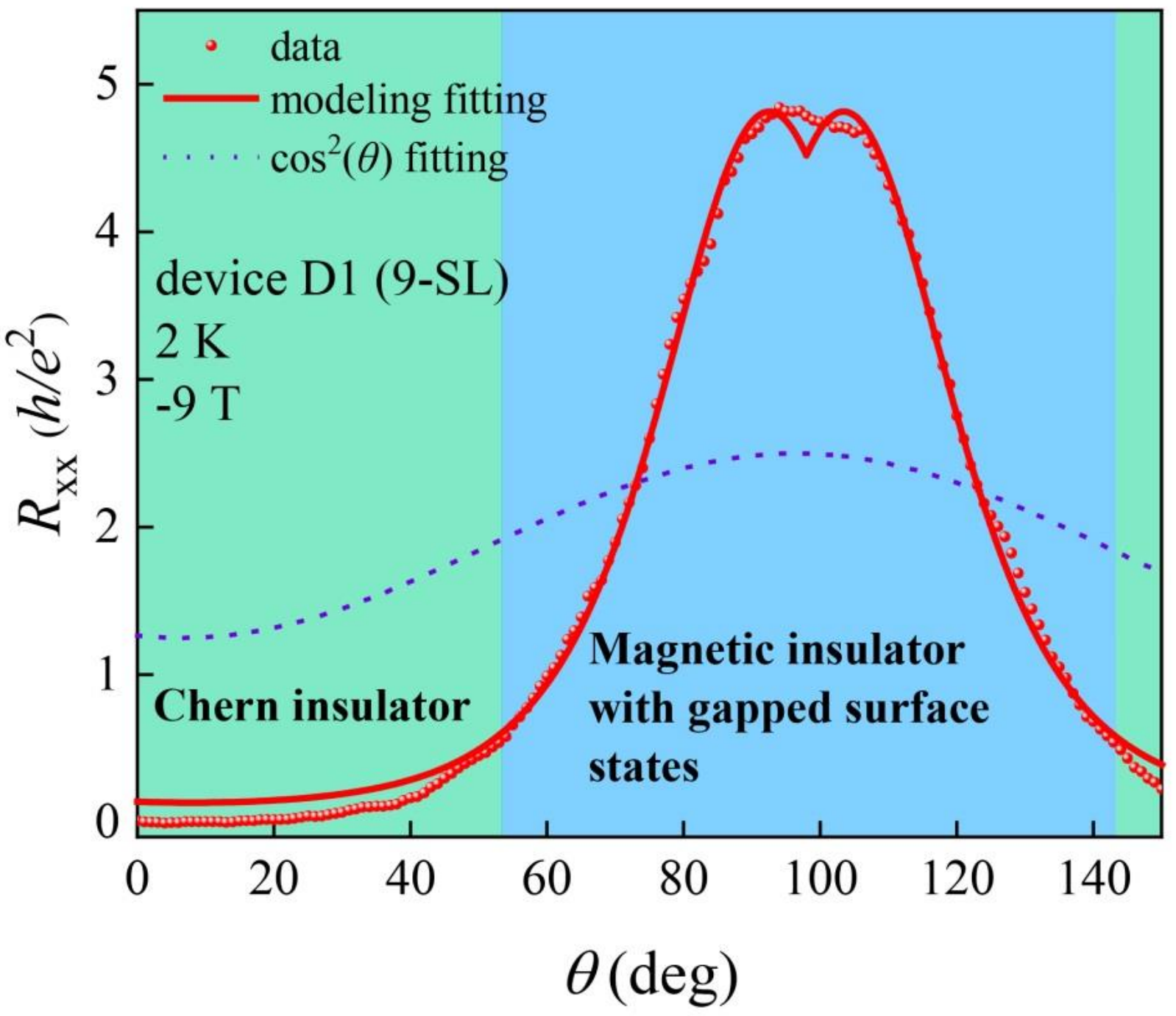

FIG. 4. Theoretical fitting of the $R_{xx}$-$\theta$ curve in $MnBi_2Te_4$ device D1 (9-SL). The red solid line is the theoretical fitting curve based on the modified Landauer-Büttiker formula, wherein the dissipative contribution from the gapped surface states in a magnetic insulator with gapped Dirac surface states is considered. The theoretical fitting shows an evident deviation in the Chern insulator (CI) region (green region, $\theta$ <52° and $\theta$ >144°). In the blue region (52°<$\theta$ <144°), the experimental data can be well fitted by the modified Landauer-Büttiker formula, indicating that the $MnBi_2Te_4$ device in this region is tuned into a magnetic insulator with gapped Dirac surface states, which suggests a field-tilt-driven topological quantum phase transition from CI to magnetic insulator with gapped Dirac surface states.

In summary, we explore the relation between magnetic structures and topological phases of $MnBi_2Te_4$ and study the evolution of the chiral edge states in $MnBi_2Te_4$ devices by angle-dependent transport measurements. As tilting the magnetic field away from the out-of-plane direction, the switch from edge dominated transport to diffusive transport is observed, evidenced by a colossal anisotropic magnetoresistance. By fitting the experimental data with a modified Landauer-Büttiker formalism, a magnetization-tuned topological quantum phase transition from CI to magnetic insulator with gapped Dirac surface states is revealed. Our work opens new avenues for identifying and manipulating novel topological quantum states.

This work is financially supported by Beijing Natural Science Foundation (Z180010), the National Key R&D Program of China (2018YFA0305600, 2017YFA0303302, 2018YFA0307100), the National Natural Science Foundation of China (11888101, 11774008, 51991340, 21975140, 12025405, 51788104, 11874035), the Strategic Priority Research Program of Chinese Academy of Sciences (XDB28000000) and the Open Research Fund Program of the State Key Laboratory of Low-Dimensional Quantum Physics, Tsinghua University (Grant No. KF202001).

J. Ge, Y. Liu. and P. Wang. contributed equally to this work.

* Corresponding author.

jianwangphysics@pku.edu.cn

Supplemental Material for

# Magnetization-tuned topological quantum phase transition in $MnBi_2Te_4$ devices

Jun Ge[1], Yanzhao Liu[1], Pinyuan Wang[1], Zhiming Xu[2], Jiaheng Li[2], Hao Li[3,4], Zihan Yan[1], Yang Wu[4,5], Yong Xu[2,6], Jian Wang[1,2,7,8*]

[1] International Center for Quantum Materials, School of Physics, Peking University, Beijing 100871, China

[2] State Key Laboratory of Low Dimensional Quantum Physics, Department of Physics, Tsinghua University, Beijing 100084, China

[3] School of Materials Science and Engineering, Tsinghua University, Beijing 100084, China

[4] Tsinghua-Foxconn Nanotechnology Research Center and Department of Physics, Tsinghua University, Beijing 100084, China

[5] Department of Mechanical Engineering, Tsinghua University, Beijing 100084, China

[6] RIKEN Center for Emergent Matter Science (CEMS), Wako, Saitama 351-0198, Japan

[7] CAS Center for Excellence in Topological Quantum Computation, University of Chinese Academy of Sciences, Beijing 100190, China

[8] Beijing Academy of Quantum Information Sciences, Beijing 100193, China

[*] Corresponding author.
jianwangphysics@pku.edu.cn

**The PDF file includes:**

Materials and Methods

Theoretical details of the modified Landauer-Büttiker formalism

Fig. S1. A schematic of the four-terminal model geometry.

Fig. S2. The evolution of the Chern insulator state as the magnetization tilts away from out-of-plane direction in device D2 (8-SL).

Fig. S3. $B$-$\theta$ phase diagram of D2.

## Materials and Methods

### Crystal growth

The high-quality single crystals of $MnBi_2Te_4$ used in this work are of the same batch as reported in our previous work [1]. Firstly, high purity $Bi_2Te_3$ and MnTe were prepared by reacting high-purity Bi (99.99%, Adamas) and Te (99.999%, Aladdin), and Mn (99.95%, Alfa Aesar) and Te (99.999%, Aladdin), respectively. Then, high-quality $MnBi_2Te_4$ single crystals were obtained by reacting the stoichiometric mixture of high purity $Bi_2Te_3$ and MnTe.

### Devices fabrication

$MnBi_2Te_4$ thin flakes were mechanically exfoliated and transferred onto 300 nm-thick $SiO_2$/Si substrates in the atmosphere. After spin-coating poly(methyl methacrylate) (PMMA) resist, standard electron beam lithography in a FEI Helios NanoLab 600i DualBeam System was carried out to define the electrodes patterns with Hall bar geometry. Metal electrodes (Cr/Au and Ti/Au, 6.5/180 nm) were deposited in a LJUHV E-400L E-Beam Evaporator after in-situ Ar plasma cleaning. Finally, the PMMA layers were removed by standard lift-off process.

### Transport measurements and magnetization measurements

Electrical transport measurements on $MnBi_2Te_4$ single crystals and devices were conducted in a 9 T-Physical Property Measurement System (PPMS-DynaCool-9, Quantum Design) and a 16 T-Physical Property Measurement System (PPMS-16T, Quantum Design), respectively. Angle-dependent magnetization measurements were carried out in a commercial magnetic property measurement system (MPMS3, Quantum Design). In the measurements of devices, Stanford Research Systems SR830 lock-in amplifiers were used to measure Hall and longitudinal resistance of the devices with an AC bias current of 100 nA at a frequency of 3.777 Hz.

### Theoretical details of the modified Landauer-Büttiker formalism

We have established a model based on the four-terminal Landauer-Büttiker formalism, $I_i = \sum_j G_{ij}(V_i - V_j)$, wherein $I_i$ and $V_i$ are the current and the voltage of the $i_{\text{th}}$ contact, and $G_{ij} = \frac{e^2}{h} T_{ij}$ is the conductance between the $i_{\text{th}}$ and $j_{\text{th}}$ contacts. The model geometry is shown in Fig. S1. Setting $V_4 = 0$, the four-terminal Landauer-Büttiker formula can be expressed as:

$$\begin{pmatrix} I \\ 0 \\ 0 \end{pmatrix} = \begin{pmatrix} G_{12} + G_{13} + G_{14} & -G_{12} & -G_{13} \\ -G_{21} & G_{21} + G_{23} + G_{24} & -G_{23} \\ -G_{31} & -G_{32} & G_{31} + G_{32} + G_{34} \end{pmatrix} \begin{pmatrix} V_1 \\ V_2 \\ V_3 \end{pmatrix}$$

$T_{ij}$ takes the form $T_{ij} = \eta \delta_{i,j-1} + t_{ij}$. Here, $\eta$ is the transmission coefficient of the unidirectional edge modes from the $i_{\text{th}}$ to the $(i+1)_{\text{th}}$ contacts, and $t_{ij}$ represents the dissipative transport. We assume that $\mathrm{t}_{23} = \mathrm{t}_{32} = \mathrm{t}_{12} = \mathrm{t}_{21} = \mathrm{t}_{34} = \mathrm{t}_{43} = \mathrm{t}$ and neglect other dissipations, then the above matrix equation can be reduced to:

$$\begin{pmatrix} I \\ 0 \\ 0 \end{pmatrix} = \begin{pmatrix} \eta + t & -\eta - t & 0 \\ -t & \eta + 2t & -\eta - t \\ 0 & -t & \eta + 2t \end{pmatrix} \begin{pmatrix} V_1 \\ V_2 \\ V_3 \end{pmatrix}$$

By solving this matrix equation, we have $V_3 = \frac{h}{e^2} \frac{t^2}{(\eta+t)^3} I$, $V_2 = \frac{h}{e^2} \frac{t(\eta+2t)}{(\eta+t)^3} I$.

From $R_{xx} = \frac{\mathrm{V_2 - V_3}}{\mathrm{I}}$, we obtain the fitting formula $R_{xx} = \frac{h}{e^2} \frac{t}{(\eta+t)^2}$ .

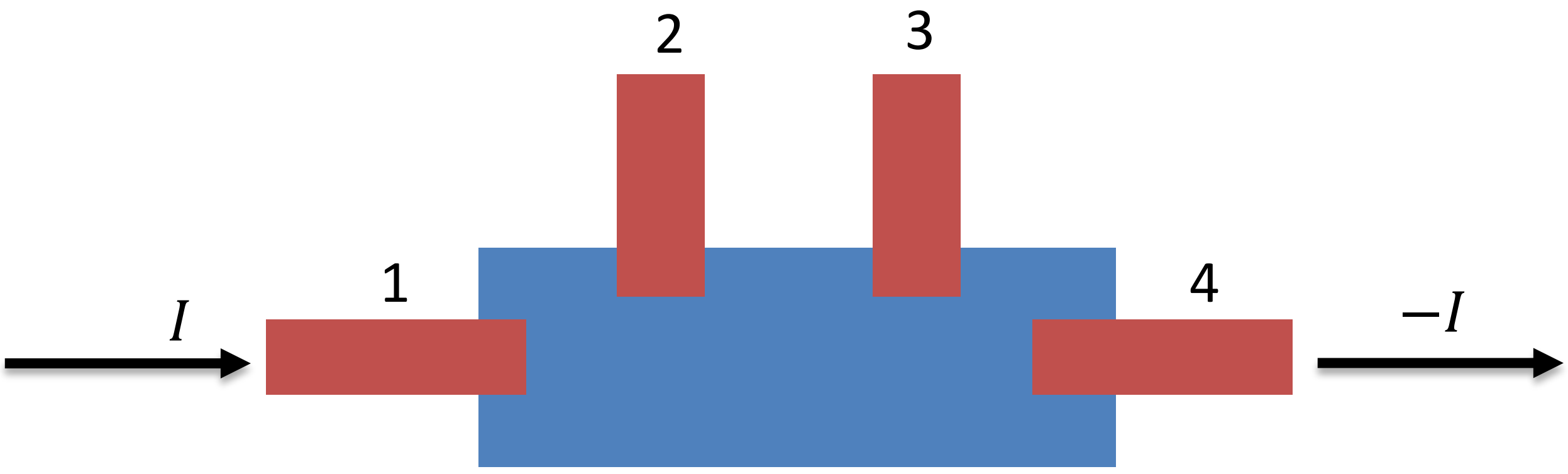

Fig. S1. A schematic of the four-terminal model geometry.

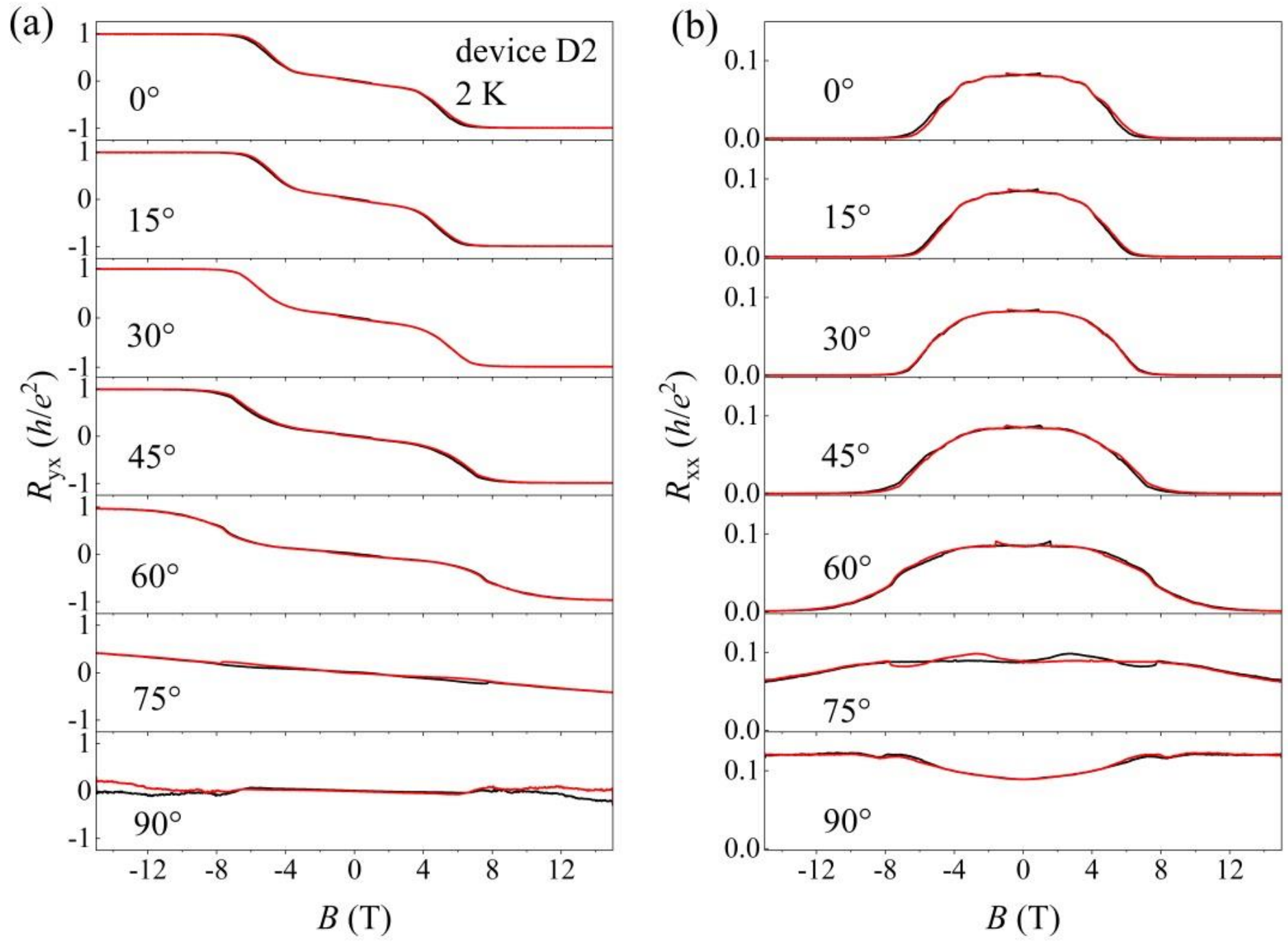

Fig. S2. (a), (b) $R_{yx}$ and $R_{xx}$ as a function of $B$ at various angles in the 8-SL device D2 ($C$=1).

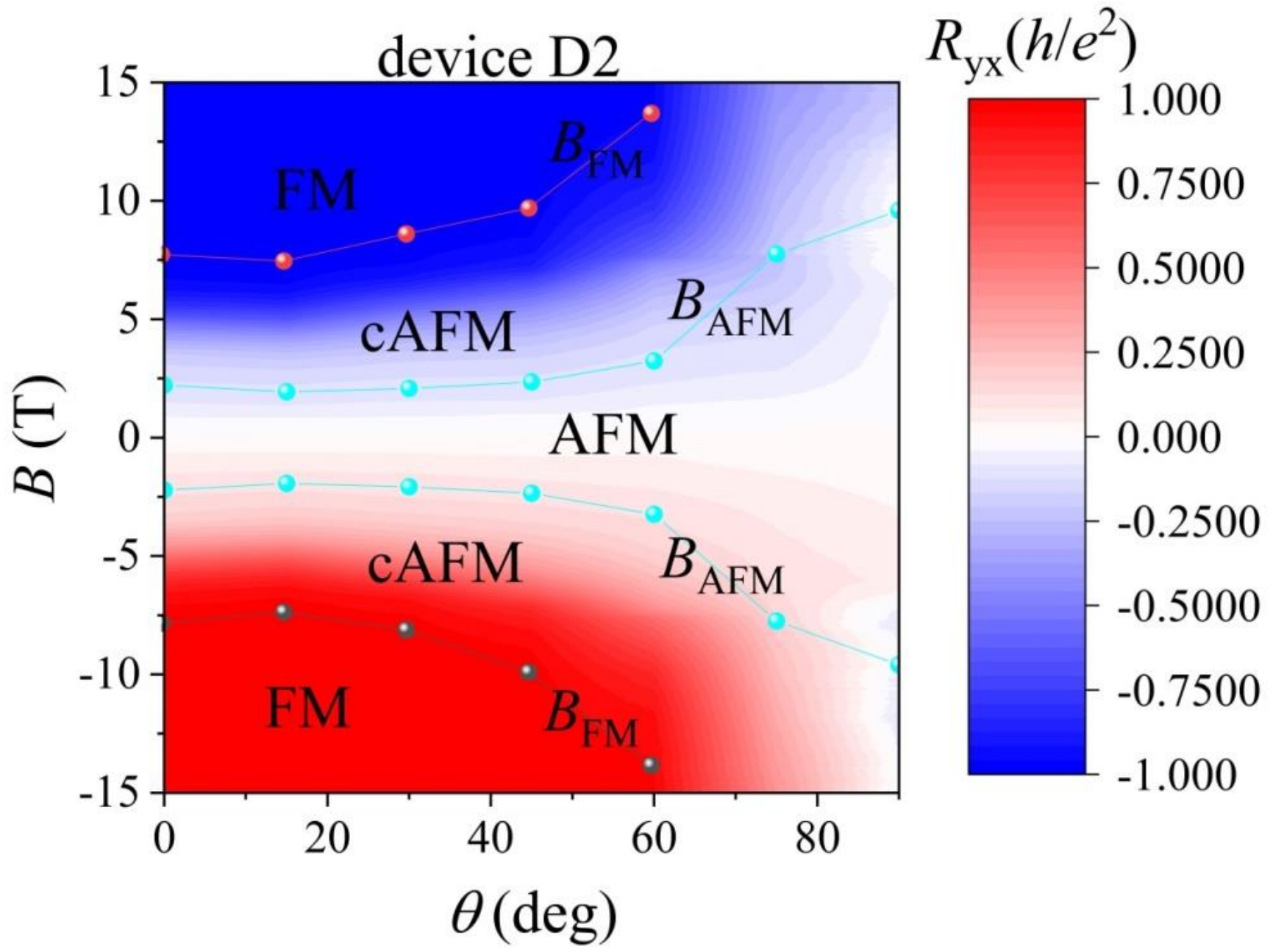

Fig. S3. $B$-$\theta$ phase diagram of the 8-SL device D2 ($C$=1).